\documentclass[a4paper,11pt]{article}
\pdfoutput=1 

\usepackage{jheppub} 

\usepackage[T1]{fontenc} 

\usepackage{graphicx}
\usepackage{epstopdf}
\usepackage{epsfig}
\usepackage{amsmath}
\usepackage{color}
\usepackage{latexsym}
\usepackage{amssymb}
\usepackage{slashed}
\usepackage{bm}
\usepackage{hyperref}
\usepackage{CJKutf8}
\usepackage{multirow}
\usepackage{booktabs}
\usepackage{verbatim}
\usepackage{epstopdf}
\usepackage{caption,subcaption}
\usepackage{lineno}
\usepackage[symbol]{footmisc}
\usepackage{hyperref}
\hypersetup{hidelinks}
\hypersetup{
colorlinks=true,
linkcolor=blue
}

\title{Probing neutrino magnetic moment at the Jinping neutrino experiment}

\author[a]{Baobiao Yue,} 
\author[a]{Jiajun Liao.\footnote{Corresponding author.}}
\author[a,b]{Jiajie Ling,\footnote{Corresponding author.}}

\affiliation[a]{School of Physics, Sun Yat-sen University, No. 135, Xingang Xi Road, Guangzhou, 510275, P. R. China}
\affiliation[b]{Key Laboratory of Particle \& Radiation Imaging (Tsinghua University), Ministry of Education, Beijing 100084, China}

\emailAdd{yuebb@mail2.sysu.edu.cn}
\emailAdd{liaojiajun@mail.sysu.edu.cn}
\emailAdd{lingjj5@mail.sysu.edu.cn}

\abstract{
 Neutrino magnetic moment ($\nu$MM) is an important property of massive neutrinos.
 The recent anomalous excess at few keV electronic recoils observed by the Xenon1T collaboration might indicate a $\sim 2.2\times10^{-11}\mu_B$ effective neutrino magnetic moment ($\mu_\nu^{eff}$) from solar neutrinos.
 Therefore, it is essential to carry out the $\nu$MM searches at a different experiment to confirm or exclude such hypothesis. 
 We study the feasibility of doing $\nu$MM measurement with 4 kton active mass at Jinping neutrino experiment using electron recoil data from both natural and artificial neutrino sources.
 The sensitivity of $\mu_\nu^{eff}$ can reach $1.2\times10^{-11}\mu_B$ at 90\% C.L. with 10-year data taking of solar neutrinos.
 Besides the intrinsic low energy background $^{14}$C in the liquid scintillator, we find the sensitivity to $\nu$MM is highly correlated with the systematic uncertainties of $pp$ and $^{85}$Kr.
 Reducing systematic uncertainties ($pp$ and $^{85}$Kr) and the intrinsic background ($^{14}$C  and $^{85}$Kr)
 can help to improve sensitivities below these levels and reach the region of astrophysical interest. 
 With a 3 mega-Curie (MCi) artificial neutrino source $^{51}$Cr installed at Jinping neutrino detector for 55 days,
 it could give us a sensitivity to the electron neutrino magnetic moment ($\mu_{\nu_e}$) with $1.1\times10^{-11}\mu_B$ at 90\% C.L..
 With the combination of those two measurements, the flavor structure of the neutrino magnetic moment can be also probed at Jinping.
 }

\begin{document} 
\maketitle
\flushbottom

\section{Introduction}
\label{sec:Intro}

Neutrino magnetic moment ($\nu$MM) \cite{Okun:1986hi,Okun:1986na,Lim:1987tk,Akhmedov:1988uk,Back:2002cd,Grimus:2002vb,Giunti:2014ixa} is related to the neutrino mass $m_\nu$ \cite{Fujikawa:1980yx,Schechter:1981hw,Kayser:1982br,Nieves:1981zt,Pal:1981rm,Shrock:1982sc}, which is verified by several neutrino oscillation experiments. Under the standard electroweak theory,
it can be expressed as
\begin{equation}
\mu_{\nu}=\frac{3 m_{e} \mathrm{G}_\mathrm{F}}{4 \pi^{2} \sqrt{2}} m_{\nu} \mu_{B} \approx 3.2 \times 10^{-19}\left(\frac{m_{\nu}}{ \mathrm{eV}}\right) \mu_{B}
\end{equation}
with $m_e$ being the electron mass, $\mathrm{G}_\mathrm{F}$ the Fermi coupling constant and $\mu_{B}=\frac{e \hbar}{4 \pi m_{e}}$ the Bohr magneton.
The current upper bound on the neutrino masses $m_\nu$ put a constraint on $\mu_\nu$ to be less than $10^{-18}\mu_B$ order. 
However, some theory extensions beyond MSM predict $\nu$MM with $10^{-(10\thicksim12)}$ $\mu_B$ order \cite{Okun:1986na, Fukugita:1987ti, Pakvasa:2003zv, Gorchtein:2006na, Bell:2006wi} for Majorana neutrino.
Moreover, some general considerations \cite{Bell:2005kz,Bell:2006dq} assert that the Dirac $\nu$MM should have a model-independent, "naturalness" upper bounds $|\mu_\nu|\lesssim10^{-14}$ $\mu_B$.
Hence if the observation of $\nu$MM is greater than $10^{-14}$ $\mu_B$ \cite{Kayser:2008Neutrino,Giunti:2008ve}, it would be an evidence of new physics and implies that neutrino might be a Majorana particle.

A previous measurement at Super-Kamiokande reported $1.1\times 10^{-10}\mu_B$ (90\% C.L.) combined with other solar neutrino and KamLAND experiments \cite{Arpesella:2008mt}.
Gemma experiment bound electron antineutrino magnetic moment ($\bar{\nu}_e$MM) below $2.9\times 10^{-11}\mu_B$ (90\% C.L.) from a reactor core with a very short baseline \cite{Beda:2013mta}.
The Borexino collaboration reported the most stringent upper limit on $\mu_\nu^{eff}$ with $2.8\times 10^{-11}\mu_B$ (90\% C.L.) \cite{Borexino:2017fbd} from solar neutrinos.

Recently, Xenon1T observed a 3 $\sigma$ event excess \cite{Aprile:2020tmw} at low electron recoil energies, which could be caused by $\nu$MM ($\mu_\nu^{eff}\in(1.4,2.8)$ $[\times 10^{-11}\mu_B]$ (90\% C.L.)) from mainly $pp$ neutrinos. 
In the meanwhile, PandaX-II sets a upper bound of $3.2\times10^{-12}\mu_B$ \cite{Zhou:2020bvf}, which does not reach the possible region of $\nu$MM in Xenon1T.
However, no current terrestrial experiments can validate the $\nu$MM that is suggested by Xenon1T.
On the other hand, for astrophysical observations, more stringent bounds are given down to $2\times10^{-12}\mu_B$ (90\% C.L.) \cite{Raffelt:1989xu,Arceo-Diaz:2015pva,Diaz:2019kim,Corsico:2014mpa}. 
Due to large uncertainties in astrophysical measurements, more precise terrestrial experiments are needed to explore $\nu$MM at $1\times10^{-11}\mu_B$ level in the near future.

A high precision measurement of solar neutrinos has been proposed by the Jinping collaboration in China, aiming to obtain high precision of solar neutrinos at the sub-percentage level \cite{JinpingNeutrinoExperimentgroup:2016nol}.
We explore the possibility of carrying the measurement of $\nu$MM at Jinping neutrino experiment via solar neutrinos. 
We also consider a MCi-scale electron capture neutrino source like mostly $^{51}$Cr \cite{Abdurashitov:2005tb,Hampel:1997fc,Abdurashitov:1998ne,Cribier:1996cq,SAGE-BEST:ConferenceTalk}, which can release sub-MeV neutrinos as well for this study, as proposed by \cite{Coloma:2014hka,Coloma:2020voz}. 

The paper is organized as follows. 
Section~\ref{sec:NaturalNv} depicts the study on $\nu$MM measurement with the natural neutrino source.
Section~\ref{sec:ArtificialNv} presents the research on $\nu$MM with a specific artificial neutrino source.
Conclusions are drawn in section~\ref{sec:Conclusions}.

\section{$\nu$MM measurement with solar neutrinos}
\label{sec:NaturalNv}
\subsection{Jinping neutrino experiment}
The Jinping neutrino experiment (Jinping) \cite{JinpingNeutrinoExperimentgroup:2016nol} aims to study MeV-scale neutrinos, including solar neutrinos, geoneutrinos and supernova neutrinos. 
It is located in one of the deepest underground laboratories in the world with 2400 m vertical rock-overburden shielding, leading to a much small cosmic-ray muon background. 
The target material is the water-based liquid scintillator (LS), whose Cherenkov light can indicate the direction of charged particles and scintillation light can be used for the precise energy reconstruction of the particles. 
The nominal energy resolution of this kind of material is nominally 500 PE/MeV. 
In this study we assume a 4 kton fiducial target mass with 5 kton total mass.
Comparing with Borexino, Jinping has a smaller cosmic-ray and a bigger detector mass. 
Therefore, Jinping could obtain more remarkable results.

 $\nu$MM is detected though the neutrino elastic scattering ($\nu$ES) from solar neutrinos or artificial neutrino source.
The cross section of $\nu$ES with $\nu$MM can be expressed as
\begin{equation}
\frac{\mathrm{d}\sigma}{\mathrm{d}T_e} (T_e,E_\nu) = 
\underbrace{\frac{\sigma_{0}}{m_{e}}\left[g_{1}^{2}+g_{2}^{2}\left(1-\frac{T_{e}}{E_{\nu}}\right)^{2}-g_{1} g_{2} \frac{m_{e} T_{e}}{E_{\nu}^{2}}\right]}_{\sigma_{\mathrm{SM}}}
+\,  
\underbrace{\pi  \frac{\alpha^2}{m_e^2} \left(\frac{\mu_\nu}{\mu_B}\right)^2\left(\frac{1}{T_{e}}-\frac{1}{E_{\nu}}\right)}_{\sigma_{\nu \mathrm{MM}}} \,.
\end{equation}
The standard model (SM) cross section contains $\sigma_0=\frac{2\mathrm{G}_\mathrm{F}^2m_e^2}{\pi} \simeq 88.06\times 10^{-46}$ $\mathrm{cm}^2$ and the electron mass $m_e=0.511$ MeV.
For $\nu_e$ and $\bar{\nu}_e$, $g_1$ and $g_2$ yields
$g_{1}^{\left(\nu_{e}\right)}=g_{2}^{\left(\bar{\nu}_{e}\right)}=\frac{1}{2}+\sin ^{2} \vartheta_{\mathrm{W}} \simeq 0.73$ and  
$g_{2}^{\left(\nu_{e}\right)}=g_{1}^{\left(\bar{\nu}_{e}\right)} = \sin ^{2} \vartheta_{\mathrm{W}} \simeq 0.23$.
Whereas for $\nu_{\mu,\tau}$ and $\bar{\nu}_{\mu,\tau}$, they obey
$g_{1}^{\left(\nu_{\mu, \tau}\right)}=g_{2}^{\left(\bar{\nu}_{\mu, \tau}\right)}=-\frac{1}{2}+\sin ^{2} \vartheta_{\mathrm{W}} \simeq-0.27$ and 
$g_{2}^{\left(\nu_{\mu, \tau}\right)}=g_{1}^{\left(\bar{\nu}_{\mu, \tau}\right)}=\sin ^{2} \vartheta_{\mathrm{W}} \simeq 0.23$.
For $\nu$MM cross section, $\mu_\nu^{eff}$ is the effective $\nu$MM in $\mu_B$ units.
The $\nu$MM cross section from neutrino magnetic moment is proportional to $(1/T_e-1/E_\nu)$ , which leads that the measurement of $\nu$MM significantly depends on the capacity of the detection at low energy. 
\subsection{Solar neutrino signals}
Jinping is far from the nuclear reactor plants. 
Therefore, the main part of the natural neutrinos are solar neutrinos at the low energy range.
The fluxes of solar neutrinos are model-dependent.
Currently, the variants of the standard solar model (SSM) can been divided into two: high metallicity (HZ) and low metallicity (LZ) \cite{Vinyoles:2016djt}.
In the following study, we assume the HZ hypothesis with $pp$ ($5.98(1\pm0.006)\times10^{10}$), $^7$Be ($4.93(1\pm0.06)\times10^9$),
$pep$ ($4.93(1\pm0.06)\times10^9$), CNO ($4.88(1\pm0.11)\times10^8$), $^8$B ($5.46(1\pm0.12)\times10^6$) and $hep$ ($7.98 (1\pm0.30)\times10^3$) in the unit of $\mathrm{cm}^{-2}\mathrm{s}^{-1}$.
In this study, the contributions from $pp$ and $^7$Be neutrinos dominate the measurement of $\nu$MM according to ref.~\cite{Borexino:2017fbd}. 
The electron recoiling signals of Carbon-Nitrogen-Oxygen (CNO) fusion circle, $pep$, $^8$B and $hep$ neutrinos can be neglected with regard to the measurement of $\nu$MM due to the small component at low energy region. 

In general, the prediction of the electron scattering signal from solar neutrinos can be counted by
\begin{equation}
\label{eq:SolarSignal}
N_{pre}(T_e) = N_e T\sum_i\phi_i\int S^{\odot}_i(E_\nu)\sum_{\alpha=e,\mu,\tau} P_{e\alpha}^i(E_\nu) \sigma_{\alpha}(E_\nu,T_e) dE\,.
\end{equation}
$N_e$ is the total electron number of the fiducial volume counted as $N_e=V\rho_{\mathrm{LS}}\rho_e \mathrm{N_A}=1.35\times10^{33}$ with the total volume $V$, $T$ is the exposure time, the LS density $\rho_{\mathrm{LS}}$, the electron density per gram $\rho_e$ (mol/g) and the Avogadro constant $\mathrm{N_A}$.
$i$ is the $i^{th}$ solar neutrino source.
$\phi_i$ is the corresponding neutrino flux.
$S^\odot_i$ is the normalized energy spectrum of such neutrino.
$P_{e\alpha}^i(E_\nu)$ is the oscillation probability, which is weighted by the different neutrino distributions in the sun, with $e-\alpha$ flavor transition from the sun to the earth.
The $i^{th}$ $P_{e\alpha}^i(E_\nu)$ can be approximately given as 
\begin{equation}
\label{eq:Pro}
P_{e\alpha}^i(E_\nu) = \int_{0}^{R_{\odot}}F_i(r)P_{e\alpha}^{\odot}(r,E_\nu)dr \,,
\end{equation}
where $R_{\odot}$ is the radius of the sun, $F_i(r)$ is the normalized $i^{th}$ neutrino distribution \cite{Bahcall:2005va} as a function of $r$, which is the distance to the center of the sun, and $P_{e\alpha}^{\odot}(r,E_\nu)$ is the oscillation probability of neutrino produced at $r$ with an energy $E_\nu$ and detected at the earth.
 As a good approximation, the matter effect from the earth is neglected in the night due to the very low neutrino energy of $pp$ and $^7$Be ($E_\nu^{max}$<1 MeV).
Therefore, $P_{e\alpha}^{\odot}(r,E_\nu)$ takes no account of the day-night effect.
$\sigma_{\alpha}$ is the cross section of $\nu$ES with $\alpha$-flavor neutrinos.

In this study, we assume 100\% detection efficiency in the fiducial volume.
\begin{figure}[htbp]
\centering
\includegraphics[width=\textwidth]{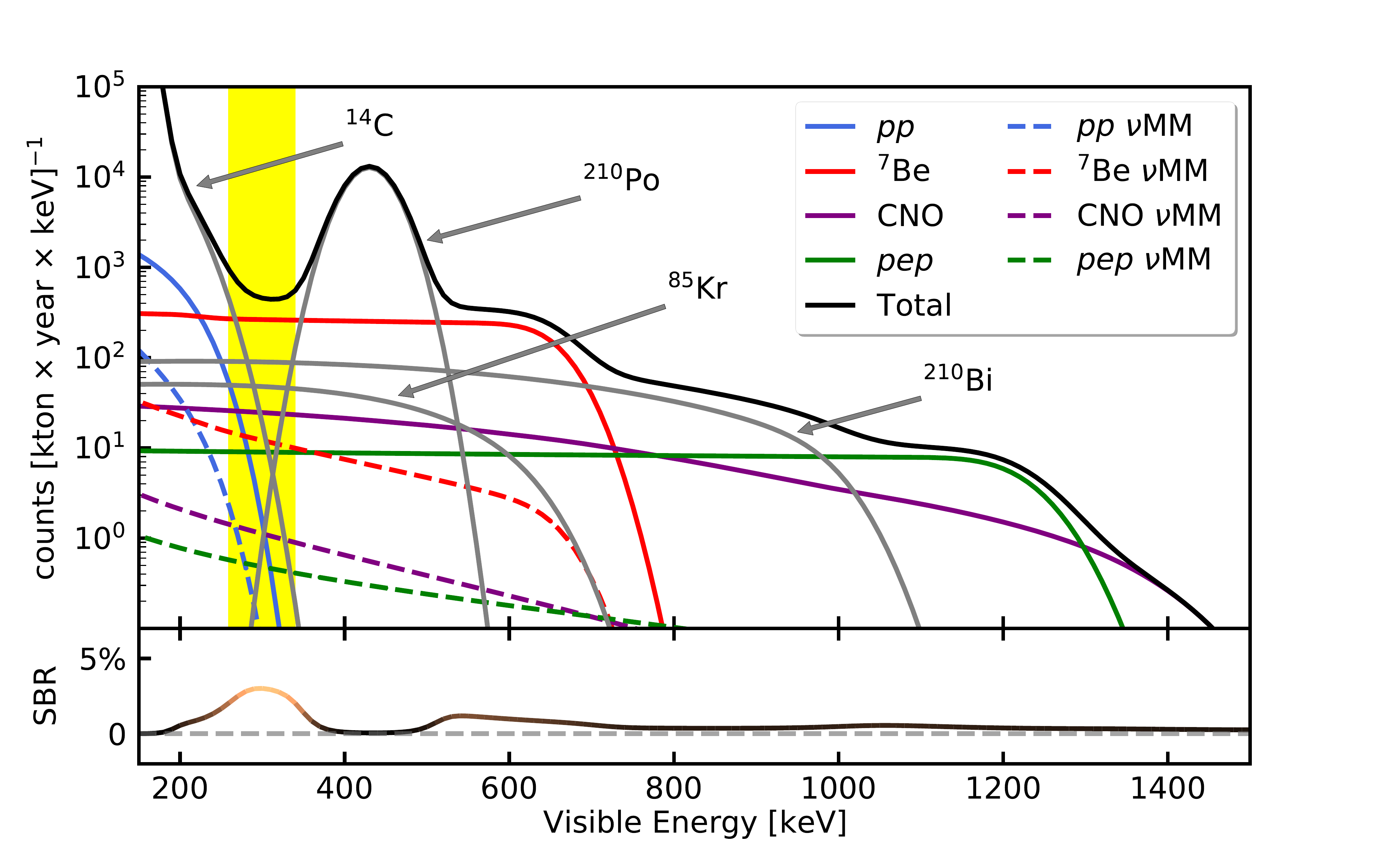}
\caption{\label{fig:1} Events distribution with signal and background components. 
The dashed lines correspond to the contributions of the magnetic moment from their sources with the same color. 
The target signal is the electronic recoil caused by $\nu$MM from all solar neutrinos. 
The SM ES signal is also background for $\nu$MM.
The yellow band represents the region with the relative large $\nu$MM signal-to-background ratio shown in the subplot for all neutrinos, assuming $\mu_\nu^{eff}=2.2\times10^{-11}\mu_B$, which is the best fit of $\nu$MM hypothesis in Xenon1T \cite{Aprile:2020tmw}.}
\end{figure}
Figure \ref{fig:1} shows the predictions of each components including signal and background. 
We generate the solar neutrino signal by using the parameters in PDG 2020 \cite{Zyla:2020zbs}.
In order to measure $\nu$MM precisely, it is crucial to evaluate accurately all components which are overwhelming the most sensitive $\nu$MM detection region, the region of interest (ROI) plotted with the yellow band in this figure.
For the solar neutrinos, the SM $\nu$ES part of $pp$ could mimic $\nu$MM ES part of itself because of the similar spectra of them.
Therefore, the external constraint of the flux of $pp$ takes an important role.
the flux of $pp$ neutrino can be mainly limited by the radiochemical constraints with $5\%$ from gallium experiments \cite{Abdurashitov:2009tn}. 
The electron recoil with $^7$Be neutrinos rises an outstanding "shoulder" structure up from about 500 to 700 keV, providing a precise measurement of $^7$Be neutrino flux.
The fluxes of CNO and $pep$ could be counted from about 900 to 1300 keV, where $\nu$MM is negligible due to the higher neutrino energy and lower fluxes than $pp$ and $^7$Be.  

\subsection{Background}
In general, three types background should be considered: cosmic-ray muon induced background, internal radioactive background and external radioactive background.
However, for $\nu$MM study, all SM $\nu$ES components are also kinds of background to $\nu$MM.
The rate of cosmic-ray muon induced background can be about 200 times lower than those in Borexino due to the depth of the Jinping laboratory \cite{JinpingNeutrinoExperimentgroup:2016nol}.
 The main low-energy background from cosmic-ray is $^{11}$C at low energy. 
 However, it does not affect the measurement of $\nu$MM due to its high energy (> 1.5 MeV).
 Other cosmogenic isotopes will not be considered in this study.
 We assume the internal radioactive background can be reduced to the same level as Borexino Phase-II \cite{Agostini:2017ixy} after purifications. 
 The intrinsic $^{14}$C of LS is $2.7\times10^{-18}g/g$ in Borexino.
The external $\gamma$-rays can be predicted with a exponential scale factor according to the distance from the edge of the fiducial volume to the detector surface.
However, it is also insignificant because it dominates at high energy range (> 1.5 MeV).
For $\nu$MM study, $^{14}$C, $^{210}$Bi, $^{85}$Kr and $^{210}$Po are the main background shown with the rate in table \ref{tab:1}. 
The pile-up of $^{14}$C-$^{14}$C has been considered for Jinping. 
We assume a naive 200 ns signal window ($\delta t$) to coarsely estimate the rate as 
\begin{equation}
R_{pile-up}\approx2\times\frac{ \left(R_{^{14}\mathrm{C}} \times M_{FV}\right) \times \left(R_{^{14}\mathrm{C}}\times M_{TV}\right) \times\delta  t}{M_{FV}}\approx 28000 \text{ cpd/kton}\,,
\end{equation}
where $M_{FV}$ is the fiducial volume mass, $M_{TV}$ is the total volume mass and $\delta t$ is the signal integral window. 
The spectrum of it is generated through the convolution with the spectrum of $^{14}$C.
Other pile-up events, mainly $^{14}$C with external $\gamma$-rays and $^{210}$Po with external $\gamma$-rays \cite{Marcocci:2020rzm}, are negligible.

\begin{table}[h!]
  \begin{center}
    \begin{tabular}{c|c}
	\hline
	\hline
      Background & Rate [cpd/kton]\\
	\hline
	\textbf{$^{14}$C} & $3.456 (1\pm 0.05)\times 10^{7}$ \\
	\textbf{$^{85}$Kr} & $68(1\pm 0.26)$ \\
	\textbf{$^{210}$Bi} & $175(1\pm 0.11)$ \\
	\textbf{$^{210}$Po} & $2600(1\pm 0.01)$ \\
      \hline
	\hline
    \end{tabular}
  \end{center}
   \caption{\label{tab:1}Background rates taken from Borexino Phase-II \cite{Agostini:2017ixy}.}
\end{table}

In general, $^{14}$C could be determined independently from the main analysis as Borexino suggested \cite{Bellini:2014uqa}.
$^{14}$C shuts down the feasibility of any signal detection at low energy due to its huge abundance below 150 keV.
Thanks to the better resolution at Jinping, $^{14}$C could not bury $pp$ and $^7$Be neutrinos severely above 150 keV.
However, the measurement of $pp$ is still challenging because of the pile-up of $^{14}$C-$^{14}$C.
Optimistically, we neglect the shape uncertainty of $^{14}$C \cite{Kuzminov:2000up}.
$^{210}$Po, overwhelming in the region of 350-550 keV, can be fitted clearly with a gaussian distribution.
In addition, $\alpha$ particle from $^{210}$Po could also be discriminated with scintillation pulse shape from $e^\pm$.
Conservatively, $^{210}$Po is still considered in this study.
$^{210}$Bi appears as a "shoulder" structure in the region from about 700 to 1000 keV with a relatively big event number.
Therefore, $^{210}$Bi could be fitted well.

Thanks to the energy resolution of Jinping experiment, there is a wide ROI, the yellow band in figure \ref{fig:1}, between $^{14}$C and $^{210}$Po spectra to measure the $\nu$MM.
However, $^{85}$Kr hides under all other components in ROI, resulting in a difficult measurement. 
In addition, It could almost freely mimic the shape and rate of $\nu$MM component, especially $\nu$MM from $^7$Be, in ROI of $\nu$MM.
That is to say, the residual $^{85}$Kr of the detector material can significantly influence the measurement of $\nu$MM. 
Therefore, the good purification and the independent measurement of $^{85}$Kr could accordingly improve the capability of $\nu$MM measurement for Jinping experiment.
In general, it is uncertain the the purification of $^{85}$Kr improves 1 or 2 order, according to the analysis in Borexino  Phase-II \cite{Agostini:2017ixy}.
Fortunately, $^{85}$Kr has a rather small branch of $\beta$ decay ($0.4\%$) with a coincidence signal, which could be selected with $18\%$ efficiency proposed by Borexino Phase-II \cite{Agostini:2017ixy}, giving a $4\%$ bound by 10-year measurement at Jinping.

\subsection{Sensitivity}
For the sensitivity study, we build a $\chi^2$ function as
\begin{equation}
\label{eq:chiSq}
\chi^2=\sum_{i}^{T_e}\frac{(N_{pre}^i-N_{obs}^i)^2}{N_{obs}^i}+\sum_{\alpha}(\frac{\delta_{\alpha}}{\sigma_{\alpha}})^2 \\,
\end{equation}
where $N_{pre}^i$ and $N_{obs}^i$ are the event number in the $i^{th}$ bin of the prediction and the observation with visible energy from 150 to 1500 keV, and $(\frac{\delta_{\alpha}}{\sigma_{\alpha}})^2$ is the penalty term to constrain solar neutrino oscillation parameters (i.e. $\theta_{12}$ and $\Delta m_{21}^2$), solar neutrino fluxes and background. 
In $(\frac{\delta_{\alpha}}{\sigma_{\alpha}})^2$, $\delta_{\alpha}$ means the difference between the value of fitting parameter $\alpha$ and the center value of its prior and $\sigma_\alpha$ is 1 $\sigma$ error.

%

\begin{figure}[htbp]
\centering
\includegraphics[width=0.93\textwidth]{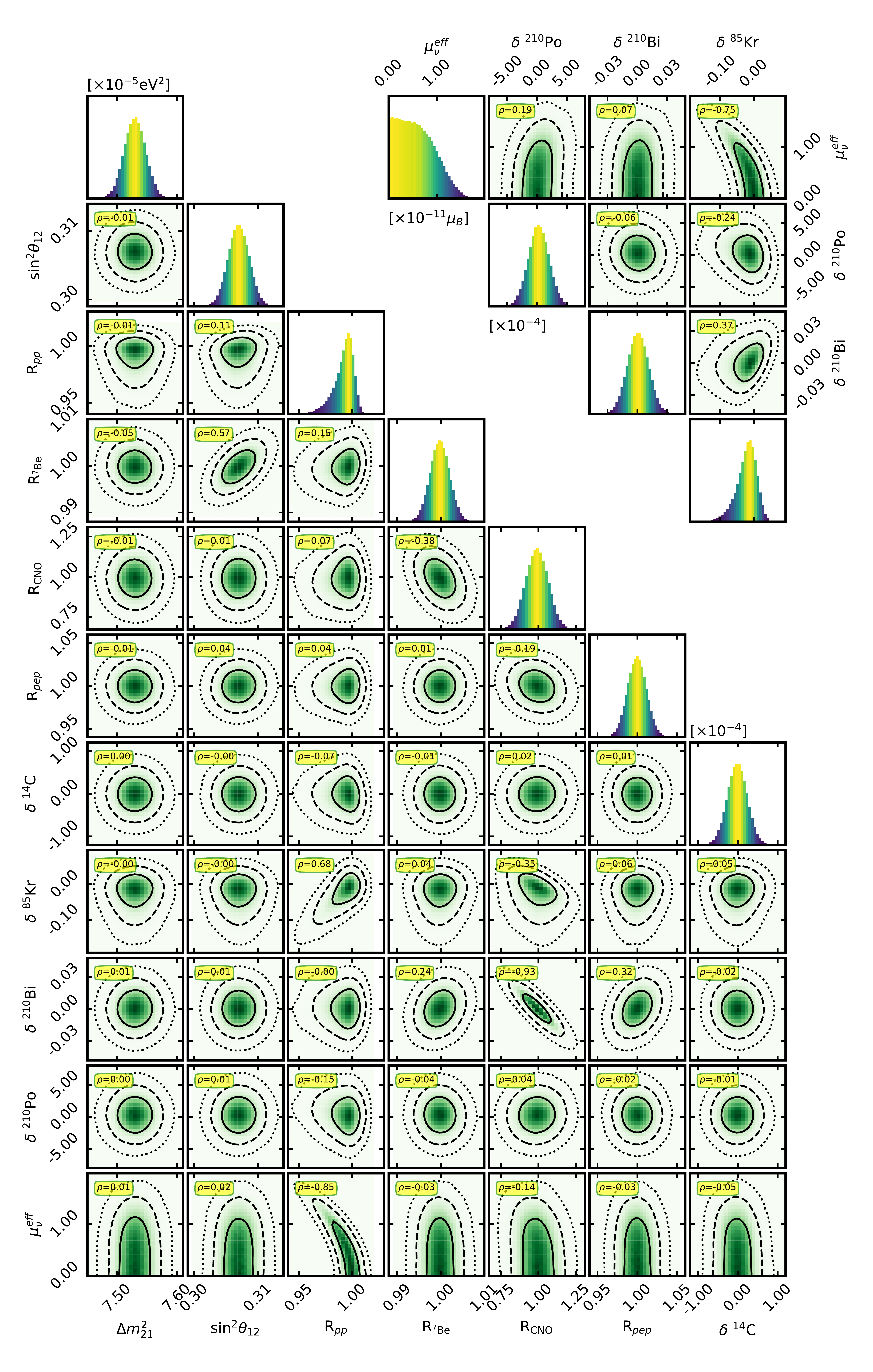} 
\caption{\label{fig:2} MCMC result with 10-year data taking. 
The solid, dashed and doted lines present 1 $\sigma$, 2 $\sigma$ and 3 $\sigma$ in all contours. 
In order to fit MCMC plot in one page, we put the lower right corner to the upper right corner.}
\end{figure}

Because there are multiple components, which are overlapping with each other at different visible energy regions, a simultaneous analysis method is necessary to take out all information for each part.
Therefore, Markov chain Monte Carlo (MCMC) technique is used to study the correlations among multi parameters, especially the relations between $\nu$MM and any other parameter, and to obtain the individual posterior distribution of each component simultaneously.
MCMC is based on a pythonic package named emcee \cite{ForemanMackey:2012ig}, 
in which a likelihood function is needed.
Therefore, we convert the $\chi^2$ function into likelihood through $\mathcal{L}=\exp{\left(-\frac{1}{2}\chi^2\right)}$ for MCMC.

Figure \ref{fig:2} presents a multi-parameter scatter plots with the full parameters by MCMC after a 10-year data taking, showing the correlations of each two parameters and the distributions of full parameters.
Solar neutrino parameters are relative to the HZ fluxes: R$_{\phi^\odot}=\phi^\odot/\phi^\odot_{truth}$. 
Background parameters are the relative differences to the background truth rates in table~\ref{tab:1}: $\delta$BG=(BG-BG$_{truth}$)/BG$_{truth}$.
This MCMC assumes that $\sin^2\theta_{12}$ and $\Delta m_{21}^2$ is constrained by JUNO, which could have accurate measurements of $\sin^2\theta_{12}$ and $\Delta m_{21}^2$ \cite{An:2015jdp} with sub-percentage 
of 0.54\% and 0.24\% respectively,
and $pp$ is constrained with 5\% by Gallium experiments. 
The MCMC scatter plot shows that $pp$ and $^{85}$Kr have the largest correlation coefficients with $\mu_\nu^{eff}$. 
Therefore, good priors of $pp$ or $^{85}$Kr do make corresponding improvements on the detection sensitivities of $\mu_\nu^{eff}$ as what we expected in the analyses of previous subsections.
From another point of view, the existence of $\nu$MM could bias the measurement of $pp$ in the similar experiments.
In figure~\ref{fig:2}, $^{210}$Bi has very strong correlations with both CNO and $pep$.
Therefore, we could fix CNO and $pep$ in the following study because $^{210}$Bi spectrum can mimic them in the low energy range.

\begin{figure}[htbp]
\centering
\includegraphics[width=0.8\textwidth]{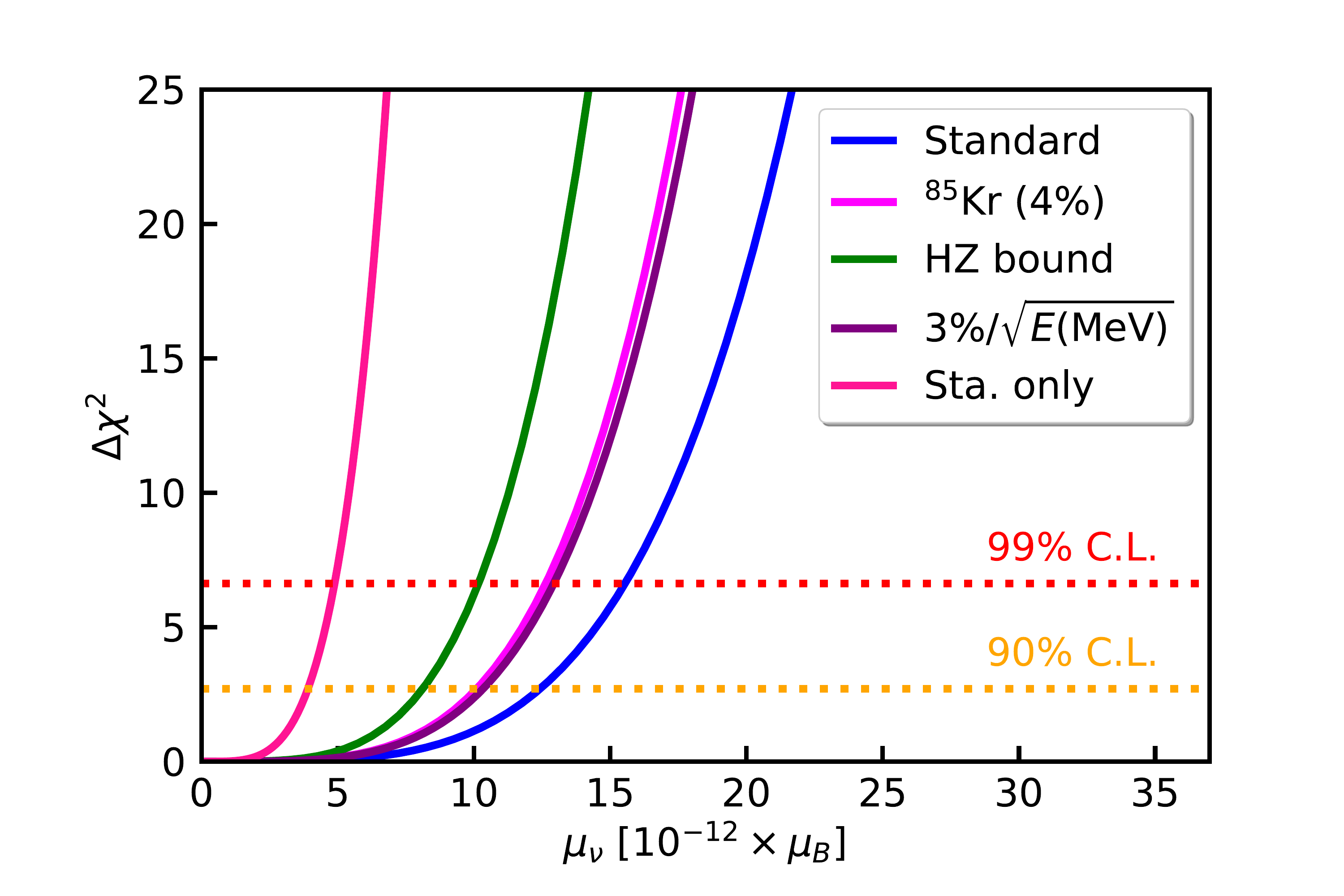}
\caption{\label{fig:3} The $90\%$ C.L. upper limit at  Jinping. 
The blue line is the standard case with only the radiochemical constraints from Gallium experiments. 
The fuchsia line presents the standard case with additional $4\%$ bound on $^{85}$Kr. 
The green line means the standard case with additional HZ model flux bounds.
The purple line is the standard case with $3\%$ energy resolution.
The deep pink line is the statistics only case.}
\end{figure}

Assuming 10-year exposure, we fix $\mu_\nu^{eff}$ by different values to obtain the $90\%$ C.L. upper limits in figure \ref{fig:3} by $\chi^2$ minimizer as another tool to study the influences with different cases. 
The sensitivities to $\nu$MM from The MCMC method and $\chi^2$ minimizer are consistent with each other.
We find that fixing $\sin^2\theta_{12}$ and $\Delta m_{21}^2$ leads no differences about the sensitivity of $\nu$MM in fitter, compared with the case that they are constrained by JUNO.
The standard case assumes that $\sin^2\theta_{12}$, $\Delta m_{21}^2$, CNO and $pep$ are fixed, $pp$ is bounded by Gallium experiments and other parameters are all free.
In figure~\ref{fig:3}, the standard case can reach $1.2\times10^{-11}\mu_B$, which is much better than other natural neutrino experiments.
The 4\% $^{85}$Kr bound can boost the $90\%$ upper limit to $1\times10^{-11}\mu_B$ as well as $3\%/\sqrt{E(\mathrm{MeV})}$, high light yield LS.
Moreover, HZ flux bound, particularly $pp$ and $^7$Be fluxes, can also improve a lot, leading to the $\nu$MM bound down to $0.8\times 10^{-11} \mu_B$.
The most ambitious sensitivity one is the statistics-only case with $3.9\times 10^{-12} \mu_B$ level at 90\% C.L. for the case of such background level in table~\ref{tab:1}.
If the background level could reduced, the result will get better.

With naive background reductions, we calculate the sensitivity to $\nu$MM with different individual reductions of $^{14}$C, $^{85}$Kr, $^{210}$Po and $^{210}$Bi.
We find that the reductions of $^{14}$C and $^{85}$Kr could significantly improve the sensitivity to $\nu$MM.
It would take a more than 10,000-fold reduction in $^{14}$C background to reach $1.0\times10^{-12}\mu_B$ level. 
A more than 1000-fold reduction in $^{85}$Kr background could at most reach $6.0\times10^{-12}\mu_B$ level.
More reduction in $^{85}$Kr could not improve any sensitivity.
Any reduction of $^{210}$Po and $^{210}$Bi could hardly improve the sensitivity to $\nu$MM.
We also find that Jinping has more than 5 $\sigma$ (3 $\sigma$) to confirm or exclude the $\nu$MM hypothesis about the recent excess in Xenon1T in 10 years (4 years).
\section{$\nu$MM with artificial neutrino source}
\label{sec:ArtificialNv}
\subsection{$^{51}$Cr neutrino signals}
\begin{figure}[htbp]
\begin{minipage}[t]{0.49\linewidth}
\centering
\includegraphics[width=1\textwidth]{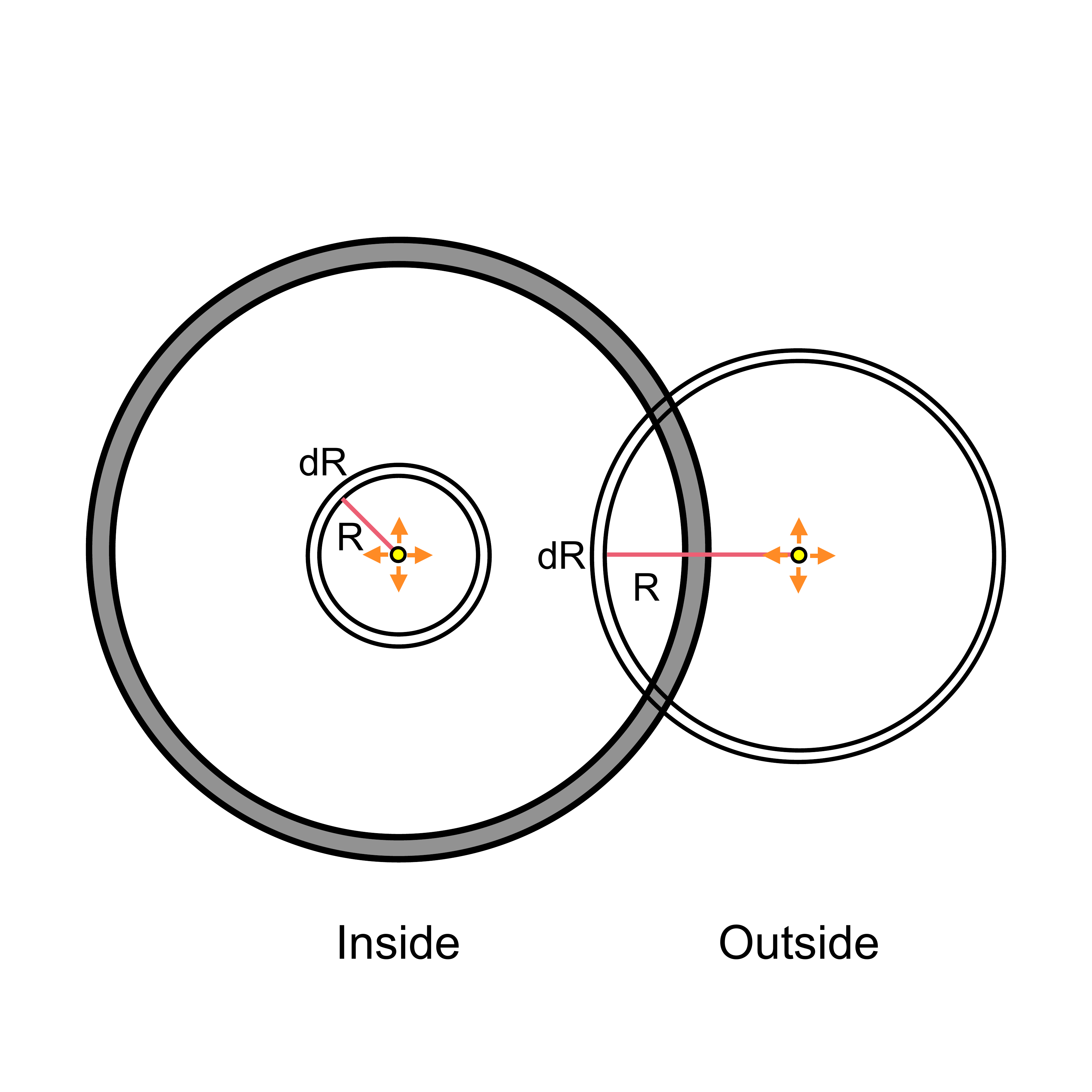}
\caption{\label{fig:4} 
The cartoon sketch about the source positions. The sizes of them are not proportional to the actual sizes.}
\end{minipage}
\hfill
 \begin{minipage}[t]{0.49\linewidth}
\centering
\includegraphics[width=1\textwidth]{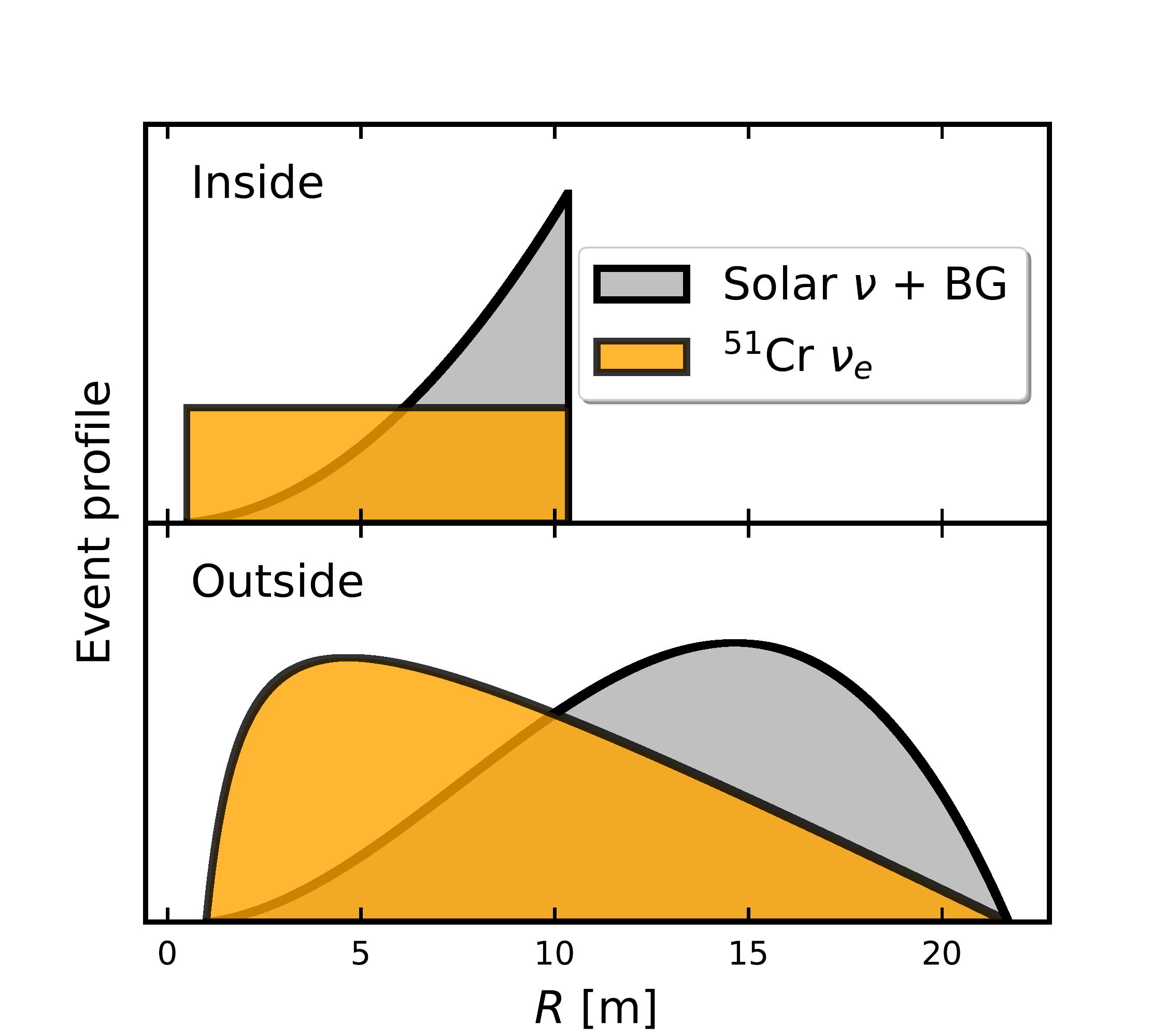}
\caption{\label{fig:5} The normalized event profile with respect to $R$ from the source.}
\end{minipage}
\end{figure}

As a specific example, a 3 MCi initialized $^{51}$Cr source \cite{Coloma:2020voz} is assumed to be placed outside with 1 meter away from the edge of the fiducial volume or inside at the center of the detector with shielding. 
Figure~\ref{fig:4} shows the cartoon sketch of the proposed source positions.
The decay of $^{51}$Cr is $^{51}\mathrm{Cr}+e^-\longrightarrow \, ^{51}\mathrm{V} +\nu_e$, with a 27.7-day half-life. The monoenergetic neutrino energies are 752 keV ($9\%$), 747 keV ($81\%$), 432 keV ($1\%$) and 427 keV ($9\%$) respectively. 

\begin{figure}[htbp]
\centering
\includegraphics[width=0.8\textwidth]{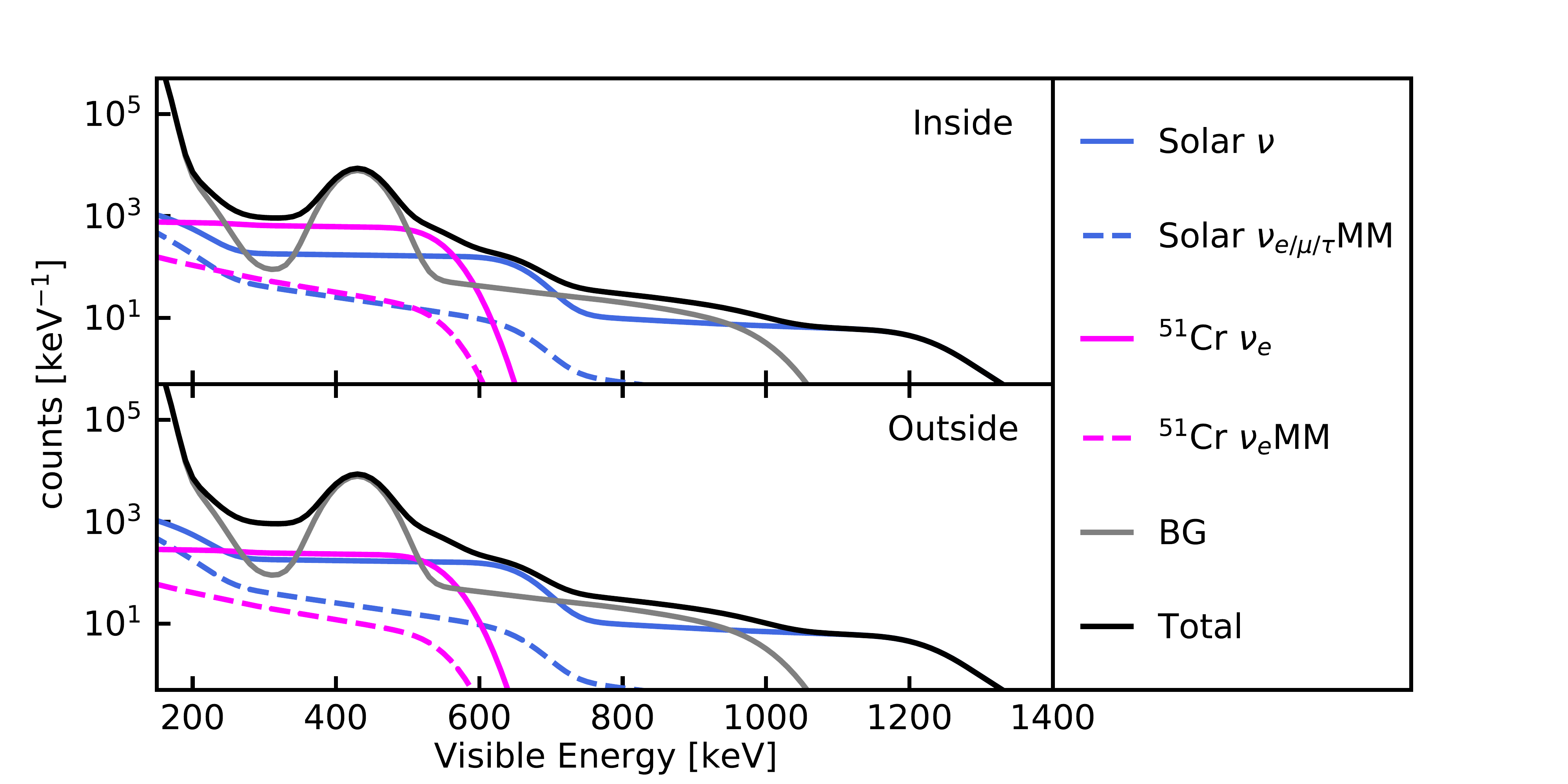}
\caption{\label{fig:6} The event number of each component by 55-day data taking with the assumption 
$(\mu_{\nu_e},\,\mu_{\nu_\mu},\, \mu_{\nu_\tau})=(3.9,\,5.8,\,5.8)$ $[\times10^{-11}\mu_B]$, the individual $90\%$ C.L. upper limits at Borexino \cite{Borexino:2017fbd}.}
\end{figure}
We calculate event vertex distribution as the function of $R$, which is the distance from the event position to the artificial source for both cases and the event number based on appendix~\ref{sec:appendixA}.
Figure \ref{fig:5} presents the event profile in the different slices as a function of $R$, assuming an ideal reconstruction performance. 
The position information can provide a strong capability to separate the signal of $^{51}$Cr and the almost uniform solar neutrinos and background in the fiducial volume.
We obtain the effective neutrino fluxes relative to the solar neutrino with 55-day $^{51}$Cr source for both cases as 
\begin{equation}
\phi_{eff}=
 \left\{
        \begin{array}{ll}
            1.27\times10^{10}\quad [\mathrm{cm}^{-2}\mathrm{s}^{-1}] \quad (\mathrm{Inside}) \\
            4.79\times10^9\quad [\mathrm{cm}^{-2}\mathrm{s}^{-1}] \quad (\mathrm{Outside})
        \end{array}
    \right.
    \,.
\end{equation}

Considering both cases, we simulate the signal and background as shown in figure \ref{fig:6}.
For solar neutrino, the event contains the $\nu$MM contribution from $\nu_e$, $\nu_\mu$ and $\nu_\tau$. 
However, the source can only contribute the $\nu_e$MM. 
Therefore, the artificial neutrino source signal can significantly break the structure of $\nu_e$, $\nu_\mu$ and $\nu_\tau$ magnetic moment in solar neutrinos, which has been presented in \cite{Borexino:2017fbd,Khan:2017djo}, when combined with solar neutrino signal. 
In this case, the probing of $\nu_e$MM will be robuster than the others.

In figure~\ref{fig:6}, $^{51}$Cr shows a outstanding "shoulder" structure between 500 and 600 keV, where the flux of $^{51}$Cr could be counted precisely.
Moreover, the amount of $\nu_e$MM shown as dash line in blue color from $^{51}$Cr source is almost equal to the summation of $\nu_e$MM, $\nu_\mu$MM and $\nu_\tau$MM shown as dash line in fuchsia color from solar neutrinos for both cases.
That is to say, the sensitivity to $\nu_e$MM should be better than other neutrino flavor magnetic moment.
The ROI of $\nu$MM with $^{51}$Cr source is the same as the solar-only case in figure~\ref{fig:1}.
\subsection{Sensitivity}
We build a similar $\chi^2$ function as eq.~\eqref{eq:chiSq} to study $\mu_{\nu_e}$, $\mu_{\nu_\mu}$, $\mu_{\nu_\tau}$ separately. 
We find that $T_e-R$ (visible energy and event vertex by $R$) two dimensional fit is almost equal $T_e-R-T$ (visible energy, event vertex by $R$ and time) three dimensional fit. 
However, $T_e$ (visible energy) one dimensional fit gets much worse sensitivity than others.
Therefore, we adopt the 2 dimensional $\chi^2$ function in the following study as
\begin{equation}
\chi^2=\sum_{i}^{T_e}\sum_{j}^{R}\frac{(N_{pre}^{i,j}-N_{obs}^{i,j})^2}{N_{obs}^{i,j}}+\sum_{\alpha}(\frac{\delta_{\alpha}}{\sigma_{\alpha}})^2 \,,
\end{equation}
where $i$ is the $i^{th}$ $T_e$ bin from 150 to 1500 keV, $j$ is the $j^{th}$ reconstructed $R$ bin. 
Compared with section~\ref{sec:NaturalNv}, the fitter has an extra parameter, the flux of $^{51}$Cr $\nu_e$.
We bound it within $1\%$ in the following analyses. 
And other parameters are consistent with the standard case in section~\ref{sec:NaturalNv}.

\begin{figure}[htbp]
\centering
\includegraphics[width=0.8\textwidth]{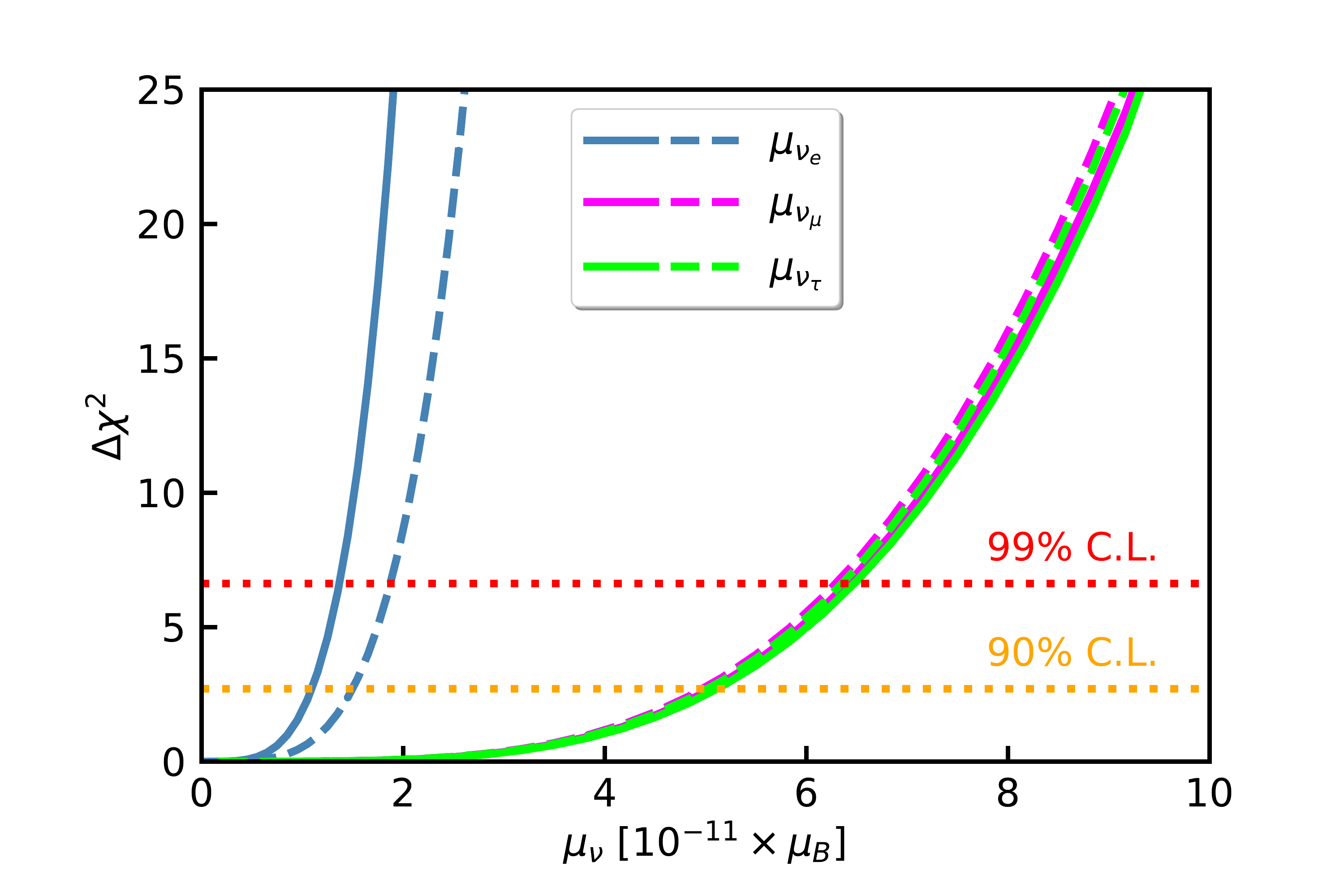}
\caption{\label{fig:7} The sensitivity with each $\nu$MM components using 55-day data taking. The solid (dash) lines represent the inside (outside) $^{51}$Cr case. }
\end{figure}

\begin{table}[htbp] 
  \centering
    \begin{tabular}{c|c|c|c}
    \hline
    \hline
     90\% C.L./$[\times10^{-11}\mu_B]$ & $\mu_{\nu_e}$ & $\mu_{\nu_\mu}$ & $\mu_{\nu_\tau}$ \cr 
    \hline
    \cline{1-4}
    Inside $^{51}$Cr & 1.1 & 5.1 & 5.1 \cr 
    \hline
     Outside $^{51}$Cr & 1.5 &  5.0 & 5.0 \cr 
    \cline{1-4}
    \hline
    \hline
    \end{tabular}
   \caption{ 90\% C.L. upper limits of $\nu$MM with 55-day data taking.
   }
  \label{tab:2}
\end{table}
 Figure \ref{fig:7} presents the individual results about $\nu$MM with different flavors. 
 Obviously, $\mu_{\nu_e}$ gets the most stringent constraint among all neutrino flavors due to the strong $\nu_e$ source. 
 When assuming the inside case, $\mu_{\nu_e}$ gets more tightly bounded than the outside case on account of more electron recoil event from the source.
 However, the bounds on $\mu_{\nu_\tau}$ and $\mu_{\nu_\tau}$ from solar neutrinos become a little weaker because $^{51}$Cr $\nu_e$ affects as a kind of background for $\mu_{\nu_\tau}$ and $\mu_{\nu_\tau}$.
  With 55-day data taking, we obtain the results with 90\% C.L. upper limits shown in table \ref{tab:2}.
 If we reduce the systematic uncertainties ($pp$ and $^{85}$Kr) and the intrinsic background ($^{14}$C and $^{85}$Kr) and enrich the strength of $^{51}$Cr source, we will obtain better results.


\subsection{Combined analyses with 55-day artificial neutrino source and 10-year solar neutrinos} 
\label{sec:Combination}
10-year data taking of solar neutrinos could give a stringent bound on $\mu_\nu^{eff}$ with the combination of $\mu_{\nu_e}$,  $\mu_{\nu_\mu}$ and  $\mu_{\nu_\tau}$. 
And 55-day data taking of $^{51}$Cr could make a tighter limit on $\mu_{\nu_e}$. 
To reach a better sensitivity, we combine them with 10-year solar neutrinos and 55-day $^{51}$Cr source neutrinos.
We define a mixing of $\mu_{\nu_\mu}$ and $\mu_{\nu_\tau}$, i.e. $\mu_{\nu_{\mu\tau}}^{eff}$, which presents the mixing of $\nu_\mu$ and $\nu_\tau$ in solar neutrinos. 
An approximate mixing $\left(\mu_{\nu_{\mu\tau}}^{eff} \right)^2 \simeq 0.49 \mu_{\nu_\mu}^2  + 0.51 \mu_{\nu_\tau}^2$ is obtained from the appendix~\ref{sec:appendixB}.

\begin{figure}[htbp]
\centering
\includegraphics[width=1\textwidth]{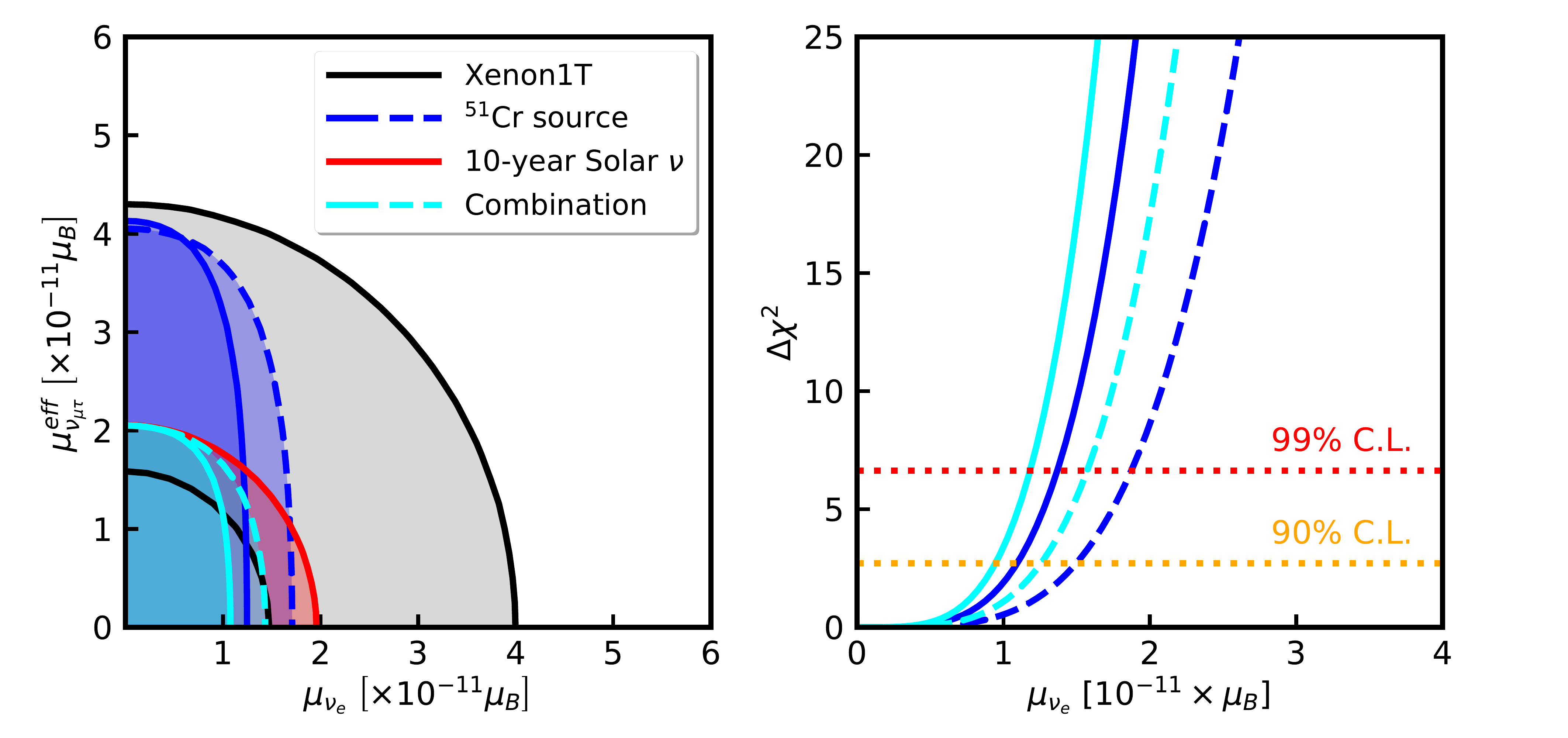}
\caption{\label{fig:8} 
The left plot is the 90\% C.L. upper limits in $\mu_{\nu_e}$ and $\mu_{\nu_{\mu\tau}}^{eff}$ plane corresponding to different cases.
And the right one is the marginalizations by $\mu_{\nu_e}$. 
The gray band is the possible area with 90\% C.L. in Xenon1T.
The solid (dash) lines represent the inside (outside) $^{51}$Cr case. 
The combination means that 10-year solar data plus 55-day $^{51}$Cr neutrino source.}
\end{figure}

The left plot of figure~\ref{fig:8} shows the 2 dimensional 90\% C.L. contours with respect to $\mu_{\nu_e}$ and $\mu_{\nu_{\mu\tau}}^{eff}$ plane. 
We reproduce the Xenon1T access with 90\% C.L. band in $\mu_{\nu_e}$ and $\mu_{\nu_{\mu\tau}}^{eff}$ plane to compare with this study.
With a $^{51}$Cr source, the constraint on $\mu_{\nu_e}$ boosts a lot. 
It is even more stringent than 10-year solar data taking.
55-day $^{51}$Cr could exclude almost all possible parameter space, especially $\nu_e$ space, in that plane.
And 10-year solar neutrino could also exclude a lot possible parameter space.
The right plot of figure~\ref{fig:8} presents the marginalization by $\mu_{\nu_e}$, showing the combination could weakly boost $\mu_{\nu_e}$ to $0.9\times10^{-11}\mu_B$ ($1.3\times10^{-11}\mu_B$) for inside case (outside case).


\section{Conclusions}
\label{sec:Conclusions}

 $\nu$MM measurement plays an important role in determining the intrinsic nature of neutrinos and probing new physics in the neutrino sector.
 The $\nu$MM has been constrained to the $3\times10^{-11}\mu_B$ level at 90\% C.L. by many terrestrial neutrino experiments.
 However, Xenon1T recently reports a hint of a $2.2\times10^{-11}\mu_B$ effective neutrino magnetic moment.
 We have calculated the feasibility of doing $\nu$MM measurement at Jinping neutrino experiment with both natural and artificial neutrino sources.

 We find the sensitivity of $\mu_\nu^{eff}$ can reach $1.2\times10^{-11}\mu_B$ level at 90\% C.L. with 10-year solar neutrino data taking at Jinping, which can validate the $\nu$MM hypothesis in Xenon1T by more than 5 $\sigma$. 
 A $4\%$ bound on $^{85}$Kr or $3\%$ energy resolution could improve the sensitivity to $1.0\times10^{-11}\mu_B$. 
 The HZ model flux bound on $pp$ and $^7$Be could lead to $0.8\times10^{-11}\mu_B$ at 90\% C.L.. 
 We find that more than 10,000-fold reduction of $^{14}$C and 1000-fold reduction of $^{85}$Kr could improve the result to the $1\times10^{-12}\mu_B$ and $0.6\times10^{-11}\mu_B$ respectively.
 The statistics-only result for the $\nu$MM is $0.4\times10^{-11}\mu_B$.

With respect to 3 MCi $^{51}$Cr neutrino source, $\nu_e$MM can be measured with $1.1\times10^{-11}\mu_B$ ($1.5\times10^{-11}\mu_B$) at 90\% C.L. for inside (outside) the detector in 55 days.
We have also considered the combination of 10-year solar neutrino and 55-day $^{51}$Cr to determine the neutrino magnetic moment induced by different neutrino flavors.
 
 \begin{figure}[htb]
\includegraphics[width=\textwidth]{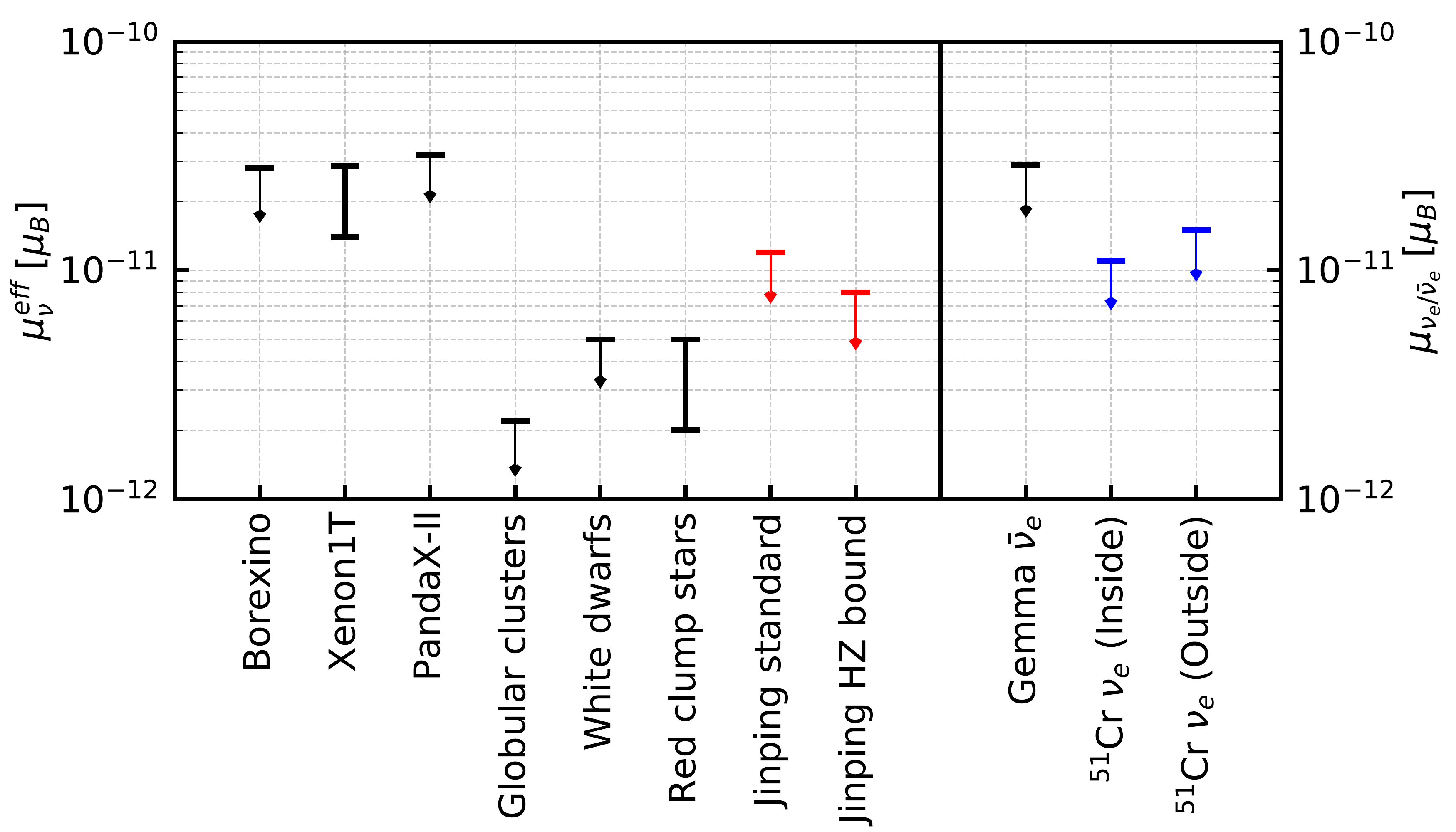}
\caption{\label{fig:9} The current status (90\% C.L.) of the neutrino magnetic moment and the sensitivities of Jinping both with solar neutrinos and artificial source. 
}
\end{figure}

 In the end, we present the current results of $\nu$MM and the sensitivities of both solar neutrino and artificial neutrino sources at Jinping in figure~\ref{fig:9}. 
 The left segment of figure~\ref{fig:9} shows $\mu_\nu^{eff}$ from this work compared to terrestrial experiments: Borexino \cite{Borexino:2017fbd}, Xenon1T \cite{Aprile:2020tmw} and PandaX-II \cite{Zhou:2020bvf}, also with the astrophysical observations: cooling of globular clusters \cite{Diaz:2019kim}, white dwarfs \cite{Corsico:2014mpa} and red clump stars \cite{Mori:2020qqd}.
The right segment presents the electronic neutrino magnetic moment from this work with $^{51}$Cr source compared to Gemma experiment \cite{Beda:2013mta}.
 Jinping could validate the $\nu$MM that is suggested by Xenon1T in the future.
 It could reach the region of astrophysical interest by reducing the systematic uncertainties and the intrinsic background or by enriching the strength of $^{51}$Cr source.
 

\section{Acknowledgments}
Jiajie Ling acknowledges the support from National Key R\&D program of China under
grant NO. 2018YFA0404013, National Natural Science Foundation of China under Grant NO. 11775315,  Key Lab of Particle \& Radiation Imaging, Ministry of Education.
Jiajun Liao acknowledges the support from the National Natural Science Foundation of China (Grant No. 11905299), Guangdong Basic and Applied Basic Research Foundation (Grant No. 2020A1515011479), the  Fundamental  Research  Funds  for  the  Central Universities, and the Sun Yat-Sen University Science Foundation.

During preparing this paper, we notice an independent and similar study [\href{https://arxiv.org/abs/2103.11771}{arXiv:2103.11771}] is also proposed and studied by Z. Ye \emph{et al.} simultaneously.
\bibliographystyle{JHEP}
\bibliography{references.bib}

\appendix
\section{Electron scattering signal distribution and the effective neutrino flux calculation for $^{51}$Cr source}
\label{sec:appendixA}
\paragraph{Outside source:}
We predict the signal event in different interfacial shell centered on the source with a radius $R$ and a thickness $\mathrm{d}R$ as

\begin{equation}
N(R,t,T_e)=N_e(R)\sum_{i} \psi(R,t,E_i)\sum_{\alpha=e,\mu,\tau} P_{e\alpha}(E_i)\sigma_\alpha(E_i,T_e) \,,
\end{equation}
 where $N_e(R)$ is the total electron number density as a function of $R$ in the shell, $\psi(R,t,E_i)$ is the neutrino flux as a function of radius $R$, time $t$ and the $i^{th}$ neutrino branch with different monoenergetic $E_i$.
 This formula can also be used for the light sterile neutrino study such as ref.~\cite{Smirnov:2020bcr}.
 In our case, there is almost no neutrino oscillation. 
 Therefore, it can reduce to 
 \begin{equation}
N(R,t,T_e)=N_e(R)\sum_{i}\psi(R,t,E_i)\sigma_e(E_i,T_e) \,,
 \end{equation}
 where, $N_e(R)=S(R)\rho_{\mathrm{LS}}\rho_e \mathrm{N_A}$ with the area of the shell $S(R)$ and $\psi(R,t,E_i)=f(E_i)\phi(R,t)=f(E_i)\frac{R_{^{51}
\mathrm{Cr}}(t)}{4\pi R^2}$, in which $f(E_i)$ is the fraction of the $i^{th}$ branch and $R_{^{51}\mathrm{Cr}}(t)$ is the decay rate initialized with 3 MCi.
 The area of the shell $S(R)$ is expressed by $S(R)=2\pi\left( 1- \frac{(R_0+x)^2+R^2-R_0^2}{2(R_0+x)R} \right)$, where $R_0$ ($\sim$10.4 m) is the radius of the fiducial volume sphere and $x$ (1 m) is the shortest distance from the source to the edge of the fiducial volume.
 So far, we obtain the total event number as a function of time $t$ and $T_e$ with the integral by $R$ from $x$ to $x+2R_0$
 
 \begin{equation}
N(t,T_e)
 = \frac{1}{2}\left[ R_0 - \frac{ (R_0+x)^2- R_0^2}{2(R_0+x)} \ln\frac{x+2R_0}{x} \right] \rho_{\mathrm{LS}}\rho_e \mathrm{N_A}  R_{^{51}\mathrm{Cr}}(t) \sum_i f(E_i)\sigma_e(E_i,T_e) \,.
 \end{equation}
 If utilizing an effective $^{51}$Cr decay rate $R^{eff}_{^{51}\mathrm{Cr}}$ during $T$ = 55 days, it reduces to 
  \begin{equation}
N(T_e)= \frac{1}{2}\left[ R_0 - \frac{ (R_0+x)^2- R_0^2}{2(R_0+x)} \ln\frac{x+2R_0}{x} \right]  \rho_{\mathrm{LS}}\rho_e \mathrm{N_A}  R^{eff}_{^{51}\mathrm{Cr}} T\sum_i f(E_i)\sigma_e(E_i,T_e) \,,
 \end{equation}
 where $R^{eff}_{^{51}\mathrm{Cr}}=6.005\times10^{16}$ Bq.
 Moreover, an effective neutrino flux $\phi_{eff}$ with respect to the whole fiducial volume can be written as $\phi_{eff}=\frac{3}{8\pi } \left[ \frac{1}{R_0^2} - \frac{ (R_0+x)^2- R_0^2}{2(R_0+x)R_0^3} \ln\frac{x+2R_0}{x}  \right] R^{eff}_{^{51}\mathrm{Cr}}$ so as to compare with the fluxes of solar neutrinos.

\paragraph{Inside source:}
With the same calculation strategy, the signal event in different spherical shell yields
\begin{equation}
N(R,t,T_e)=\rho_{\mathrm{LS}}\rho_e\mathrm{N_A}R_{^{51}\mathrm{Cr}}(t)\sum_i f(E_i)\sigma_e(E_i,T_e) \,,
\end{equation}
which is an uniform distribution of $R$.
With an integral by $R$ and $t$, we obtain
\begin{equation}
N(T_e) = \rho_{\mathrm{LS}}\rho_e\mathrm{N_A}R_{^{51}\mathrm{Cr}}^{eff}(R_0-r)T\sum_i f(E_i)\sigma_e(E_i,T_e) \,,
\end{equation}
where $r$ is the radius (0.5 m) of the sphere source shielding.
In addition, we obtain an effective neutrino flux $\phi_{eff}=\frac{3}{4\pi (R_0^2+rR_0+r^2)}R^{eff}_{^{51}\mathrm{Cr}}$ to compare with the solar neutrino fluxes.

\section{An effective neutrino magnetic moment mixing $\mu^{eff}_{\nu\tau}$}
\label{sec:appendixB}

We define an effective mixing of $\mu_{\nu_\mu}$ and $\mu_{\nu_\tau}$, $\mu_{\nu_{\mu\tau}}^{eff}$.
Therefore, \eqref{eq:SolarSignal} can be modified as 
\begin{equation}
N_{pre} =N_e\sum_i\phi_i\int \int S^{\odot}_i(E)\left(P_{ee}^i(E) \sigma_{e}(E,T_e) + (P_{e\mu}^i(E)+ P_{e\tau}^i(E))\sigma_{\mu\tau}(E,T_e) \right)dE dT_e \,,
\end{equation}
with $\sigma_{\mu\tau}=\sigma_{\mu\tau}^{\mathrm{SM}}+\sigma_{\mu\tau}^{\nu_{\mu\tau}\mathrm{MM}}$,
and $\sigma_{\mu\tau}^{\nu_{\mu\tau}\mathrm{MM}}= \pi  \frac{\alpha^2}{m_e^2} \left(\frac{\mu_{\nu_{\mu\tau}}^{eff}}{\mu_B}\right)^2  \left(\frac{1}{T_{e}}-\frac{1}{E_{\nu}}\right)$.
Moreover, remove the SM and $\nu_e$MM contributions from the event number, resulting in
\begin{equation}
\label{eq:EffNuMM1}
\begin{aligned}
&\sum_i\phi_i\int \int \pi r_0^2 S^{\odot}_i(E)
\left(P_{e\mu}^i(E) 
 + P_{e\tau}^i(E) \right)(\mu_{\nu_{\mu\tau}}^{eff})^2 \left(\frac{1}{T_{e}}-\frac{1}{E_{\nu}}\right)dE dT_e
\\
=
&\sum_i\phi_i\int \int \pi r_0^2 S^{\odot}_i(E)
\left(P_{e\mu}^i(E) \mu_{\nu_\mu}^2
 + P_{e\tau}^i(E) \mu_{\nu_\tau}^2 \right)\left(\frac{1}{T_{e}}-\frac{1}{E_{\nu}}\right)dE dT_e
\,,
\end{aligned}
\end{equation}
where $\mu^{eff}_{\nu\tau}$ can be split into $\mu_{\nu_\mu}$ and $\mu_{\nu_\tau}$. 
In this study, $^7$Be (862 keV) dominates the proportions of $\mu_{\nu_\mu}$ and $\mu_{\nu_\tau}$ in $\mu_{\nu_{\mu\tau}}^{eff}$ at ROI, the yellow band in figure~\ref{fig:1}.
Therefore, \eqref{eq:EffNuMM1} can reduce to 
\begin{equation}
\label{eq:EffNuMM2}
\left(P_{e\mu}^{^7\mathrm{Be}}
 + P_{e\tau}^{^7\mathrm{Be}} \right)\left(\mu_{\nu_{\mu\tau}}^{eff}\right)^2 
\simeq
 P_{e\mu}^{^7\mathrm{Be}} \mu_{\nu_\mu}^2
 + P_{e\tau}^{^7\mathrm{Be}} \mu_{\nu_\tau}^2 
\,,
\end{equation}
 where $P_{e\alpha}^{^7\mathrm{Be}}$ obeys \eqref{eq:Pro}. 
 Further more, the average oscillation probability $P_{e\alpha}^{^7\mathrm{Be}}$ approximates $P_{e\alpha}$($r=0.06R_{\odot}$, $E=862$ keV), the probability from the densest point of $^7$Be production in the sun to the earth.
 Therefore, we get $\left(\mu_{\nu_{\mu\tau}}^{eff} \right)^2 \simeq 0.49 \mu_{\nu_\mu}^2  + 0.51 \mu_{\nu_\tau}^2$ with the neutrino oscillation parameters in ref. \cite{Zyla:2020zbs}.

\end{document}